\documentclass[9pt]{article}
\usepackage{subeqnarray,subfigure}
\usepackage{dirtytalk}
\usepackage{braket}
\usepackage{csquotes}
\usepackage{epsfig,amsmath,amssymb,mathtools}
\usepackage{enumitem}
\usepackage{mathrsfs,authblk}
\usepackage{tikz,lipsum,lmodern}
\usepackage{empheq}
\usepackage{soul,ulem}
\usepackage{setspace}
\usepackage{verbatim}
\usepackage[pagebackref=true, colorlinks=true]{hyperref}
\definecolor{redish}{rgb}{0.7,0.2,0.0}  
\definecolor{bluish}{rgb}{0.2,0.5,0.8}
\hypersetup{linkcolor=redish,          
	citecolor=blue,        
	filecolor=magenta,      
	urlcolor=bluish}          
\usepackage{setspace}
\title{Wigner distribution of Sine Gordon and Kink solitons }
\author[1]{Ramkumar Radhakrishnan 
	 \footnote{rradhak2@ncsu.edu}}
\author[2]{Vikash Kumar Ojha 
	  \footnote{vko@phy.svnit.ac.in}}
\affil[1]{{\small Department of Physics, North Carolina State University, \linebreak Raleigh\\, USA.  }}
\affil[2]{{\small Department of Physics,\linebreak Sardar Vallabhbhai National Institute of Technology, \linebreak Surat 395007, Gujarat, India.  }}
\date{~}               
\begin{document}
\maketitle
\begin{abstract}
Wigner distributions play a significant role in formulating the phase space analogue of quantum mechanics. The Schrodinger wave-functional for solitons is needed to derive it for solitons. The Wigner distribution derived can further be used for calculating the charge distributions, current densities and wave function amplitude in position or momentum space. It can be also used to calculate the upper bound of the quantum speed limit time. We derive and analyze the Wigner distributions for Kink and Sine-Gordon  solitons by evaluating the Schrodinger wave-functional for both solitons. The charge, current density, and quantum speed  limit for solitons are also discussed which we obtain from the derived analytical expression of Wigner distributions.
\end{abstract}
\section{Introduction}
One of the seminal works in semi-classical physics is carried out by Wigner, who combines the distribution of a particle's position (coordinate) and momentum in terms of a wave function. This function which is known as the Wigner function or Wigner distribution\cite{wigner} shows the phase space formulation of quantum mechanics. This function which is defined by Wigner is not unique, and there are more such functions generally known as the quasi-probability distribution function. The Wigner distribution is similar to the probability distribution function but do not satisfy all the axioms required to call them probability distribution function. For example, they are not always positive definite and normalized to 1. However, it acts as a standard tool to study the quantum-classical interface \cite{qmcm}. The classical particle is represented by a point with its position and momentum as coordinates in the phase space \cite{cm}. For a given ensemble of particles, the Liouville density \cite{qmwigner} gives the probability distribution of that ensemble. The probability distribution is of sheer importance as it helps find the trajectory of that entire ensemble. In the case of an ensemble of quantum particles, an equal representation is not possible due to the uncertainty principle \cite{uncertain}. Therefore, Wigner distribution, mentioned above, comes to the rescue as it gives a quasi-probability distribution for that ensemble. However, it does not satisfy all the properties of conventional classical probability distribution and becomes negative \cite{negative} in some regions of the phase space. Thus Wigner distribution helps study the quantum analogue of the classical phase space approach. It has a broad range of applications namely in the field of quantum optics \cite{optics,optics2,optics3,optics4,optics5}, quantum computing \cite{qic, qic2,computing1,computing2,computing3}, signal processing \cite{sp,signal1,signal2,signal3,signal4,signal5}, quantum chromodynamics \cite{qcd1,chromo1,chromo2,chromo3}, etc. \vspace{1em}\\ 
In this work, we have calculated the Wigner distribution (one of the several quasi-probability distributions defined in the literature) of the classical solitons namely, the kink and the Sine-Gordon solitons. Solitons are the solutions of the classical field equations, which are similar to particles, sometimes referred to as pseudo particles\cite{raja}. We have also calculated the charge distributions \footnote{ We notice
that the words charge or current densities can be misleading
when a wave packet is partially transmitted or reflected, and
the charge is in fact either fully transmitted or fully reflected,
not both, when measured.} and current densities of the same, using the results obtained from the calculation of the Wigner distribution. The motivation of us to calculate the Wigner distribution of these solitons is due to the behaviour of these solitons as described in \cite{raja}, which says that soliton forms the solution of semi-classical approximation in the second quantized relativistic field theory. Moreover, it also emphasises a similar idea as a quasi-probability distribution. The study of soliton is restricted to particle physics but has a widespread application in condensed matter physics and quantum computing. Therefore, we have given a short glimpse of how the Wigner distribution helps us to find the Classical speed limit time \cite{qsl}, semi-classical speed limit time \cite{qsl}, and Quantum speed limit time \cite{deffner} in the present context of soliton towards the end, as the quantum speed limit time \cite{deffner} forms the foundations of quantum information theory.\vspace{1em}\\
The article is organized as follows. In section (\ref{section:2}) we derive and discuss the Wigner distribution, charge, and current density of the Sine-Gordon soliton using the Schrodinger wave-functional. Section (\ref{section:3}) deals with the Wigner distribution, charge, and current density of the Kink soliton. Finally, we conclude in section (\ref{section:4}) with our results, while section (\ref{section:5}) is dedicated to the acknowledgements.
\section{Wigner distribution of Sine-Gordon soliton} \label{section:2}
\begin{figure}[hbt]
    \centering
    \includegraphics[scale=0.8]{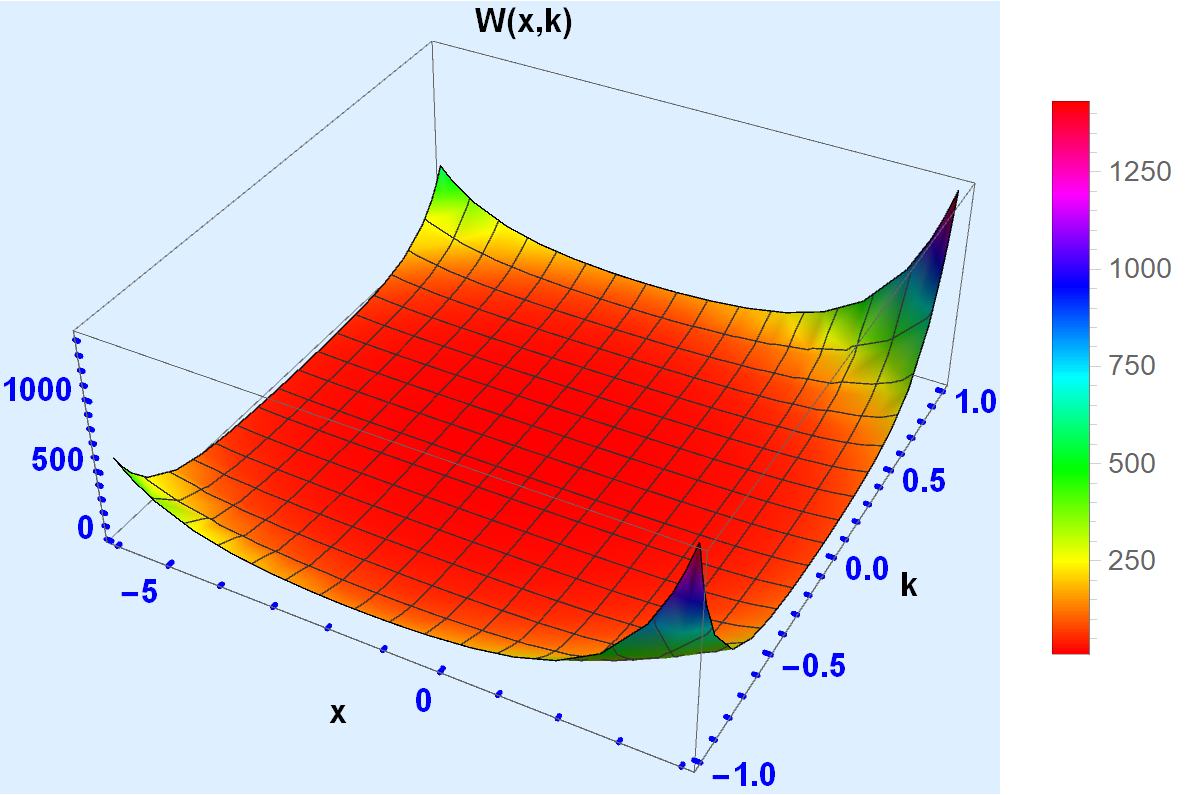}
   \caption{\label{WDsineGordon} 3D plot of Wigner distribution of Sine - Gordon soliton for $k_{0}\rightarrow 1, \hbar\rightarrow 1, n_{k} \rightarrow 0, n_{-k} \rightarrow 0, m\rightarrow 1.$}
\end{figure}

One of the quite common methods to represent a quantum mechanical system in phase space is by Wigner representation.  We review the method of shifted hamiltonian to derive the wave functional, which is prescribed in \cite{main}. The general notion of the Wigner distribution $(\mathcal{W}(x,k))$ for a given state $\psi$ can be given by
\begin{equation}\label{eq:1}
     \mathcal{W}(x,k) = \frac{1}{\pi\hbar}\int^{\infty}_{-\infty} dy \psi^{*}(x+y)\psi(x-y)e^{\frac{2iky}{\hbar}}.
\end{equation}
Let us consider a  simple classical soliton \cite{soliton} which is a solution of non-linear hyperbolic differential equation that is the Euler-Lagrange equation of the following Lagrangian density,
 $$\mathcal{L}_{SG} = \frac{1}{2}\partial^{\mu}\phi \partial_{\mu}\phi- V(\phi),$$
where $V(\phi)$ represents the potential energy given by
$$V(\phi)= \frac{-m^{2}}{\lambda} (\mathrm{cos}(\sqrt{\lambda} \phi) - 1),$$
where $m$ is the mass parameter, $\lambda$ is the coupling constant and $\pi$ is the conjugate to the field $\phi$. The Hamiltonian density of the system can be defined as
\begin{equation}
    \mathcal{H} = \frac{1}{2}:[\pi(x)]^{2}:+\frac{1}{2}:(\partial_{x}\phi(x))^{2}:-\frac{m^{2}}{\lambda} :(\mathrm{cos}(\sqrt{\lambda} \phi) - 1):,
\end{equation}
and the Hamiltonian of this system can be given by
\begin{equation}
    H = \int dx \mathcal{H} = \int dx \Bigg[\frac{1}{2}:[\pi(x)]^{2}:+\frac{1}{2}:(\partial_{x}\phi(x))^{2}:-\frac{m^{2}}{\lambda} :(\mathrm{cos}(\sqrt{\lambda} \phi) - 1):\bigg].
\end{equation}
The classical solution of the Sine-Gordon soliton is
\begin{equation}
\label{eq:25}
\phi^{cl}_{SG}(x)=\frac{4}{\sqrt{\lambda}}\mathrm{tan}^{-1}[e^{mx}].
\end{equation}
The above function does not satisfy all the properties of a well-behaved quantum wave function and hence can not be used to  evaluate the Wigner distribution. However, it can be used to derive the Schrodinger wave-functional by following the procedure of the recent article \cite{main}. The Schrodinger wave-functional is a well-behaved wave-function and can be used to derive the Wigner distribution. We start reviewing the quantisation of Sine Gordian soliton in its oscillator modes \cite{qft}. 
\begin{equation}
  \phi(x) = \int \frac{dk}{2\pi}\phi_{k}e^{-ikx},\ \phi_{k}=\frac{1}{\sqrt{2k_{p}}}(a^{\dagger}_{k}+a_{-k}) 
\end{equation}
\begin{equation}
 \pi(x) = \int \frac{dk}{2\pi}\pi_{k}e^{-ikx} ,\ \pi_{k}=i\sqrt{\frac{k_{p}}{2}}(a^{\dagger}_{k}-a_{-k})    
\end{equation}
where $k_{p}=\sqrt{m^{2}+p^{2}}$, $[a_{r}, a^{\dagger}_{s}] = 2\pi\delta(r-s)$, $[\phi_{r},\pi_{s}] = 2\pi i\delta(r+s)$. We follow the technique adopted in \cite{main} by defining a new Hamiltonian $H^{\prime}$ which is related to the original Hamiltonian through a similarity transformation. We define the new Hamiltonian $H^{\prime}$ as follows
\begin{equation}
    H^{\prime} = X^{-1} H X ,\ X=exp\bigg(-i\int dx \phi^{cl}\pi(x)\bigg),
\end{equation}
where $X$ is the translation operator which transforms the solution of the soliton. $H^{\prime}$ describes the oscillations in the soliton configuration. Since the Hamiltonian is normal ordered, regularization is not required. The transformed Hamiltonian $H^{\prime} = \mathcal{F}_{0}+\int dx \mathcal{H}^{\prime}$ where $\mathcal{F}_{0}$ represents the classical soliton energy and $H^{\prime}$ represents the quantised soliton energy. 
\begin{equation} \label{modifiedH}
    \mathcal{H}^{\prime} = \frac{:\pi(x)^{2}:}{2} + \frac{:(\partial_{x}\phi(x))^{2}}{2}:+\bigg(\frac{m^{2}}{2}-m^{2}sech^{2}(mx)\bigg ):\phi^{2}(x): 
\end{equation}
The classical equations of motion have constant frequency solutions $h_{k}(x)$ which is parameterized by $k$ and bound state solution $h_{B}(x)$ representing the soliton Goldstone mode
\begin{equation}
h_{k}(x)= \frac{e^{-ikx}}{\sqrt{1+\frac{m^{2}}{k^{2}}}}\bigg(1-i\frac{m}{k}tanh(mx)\bigg ) ,\ h_{B}(x) = \sqrt{\frac{m}{2}}sech(mx)
\end{equation}
with their respective frequencies
\begin{equation}
\omega_{k}=\sqrt{m^{2}+k^{2}} ,\ \omega_{B}= 0.
\end{equation}
By the property of normalization and orthogonality, we get
\begin{equation} \label{orthogonal}
\int dx h_{k_{1}}(x)h^{*}_{k_{2}}(x) = 2\pi\delta(k_{1}-k_{2}) ,\  \int dx|h_{B}(x)|^{2} = 1,
\end{equation}
they also satisfy the reality conditions $h^{*}_{k}(x) = h_{-k}(x),\ h^{*}_{B}(x)= h_{B}(x) $. It was convenient to decompose the field $\phi(x)$ to plane waves to obtain the Heisenberg operators $a_{k}$ in the ground state sector. Here we will be able to decompose the field $\phi(x)$ into the constant frequency solutions.
\begin{equation}
 \phi(x)= \phi_{0}h_{B}(x)+ \int \frac{dk}{2\pi} \phi_{k}h_{k}(x),\ \pi(x) = \pi_{0}h_{B}(x)+\int \frac{dk}{2\pi}\pi_{k}h_{k}(x)
 \end{equation}
 \begin{equation}
  \phi_{k} = \frac{1}{\sqrt{2w_{k}}}(b^{\dagger}_{k}+b_{k}),\  \pi_{k}=i\sqrt{\frac{w_{k}}{2}}(b^{\dagger}_{k}-b_{-k}).     
\end{equation}
Using the completeness relations, eq. \eqref{orthogonal} can be inverted as follows
\begin{equation}
    \phi_{k} = \int dx \pi(x)h^{*}_{k}(x) ,\ b^{\dagger}_{k} = \sqrt{\frac{\omega_{k}}{2}}\phi_{k} - \frac{i}{\sqrt{2\omega_{k}}}\pi_{k} ,\  b_{-k} = \sqrt{\frac{\omega_{k}}{2}}\phi_{k} + \frac{i}{\sqrt{2\omega_{k}}}\pi_{k},
\end{equation}
and
\begin{equation}
\phi_{0} = \int dx \phi(x)h_{B}^{*}(x) ,\ \pi_{0}=\int dx \pi(x)h_{B}^{*}(x) ,\ \phi_{k}=\int dx \phi(x)h_{k}^{*}(x).
\end{equation}
The commutation relations can be fixed with the usual norms. We get the answer for the transformed Hamiltonian as $H^{\prime}$, as given in \cite{main2}
\begin{equation}
    H^{\prime} = F_{1} + \int \frac{dk}{2\pi} \omega_{k}b^{\dagger}_{k}b_{k} + \frac{\pi_{0}^{2}}{2},
\end{equation}
where $F_{1}$ is the one loop soliton energy. Let us have a soliton ground state be $\ket{k}$ and we define an operator $\hat{A}$ then
\begin{equation}
    \ket{k} = X\hat{A}\ket{0}.
\end{equation}
For $E$ to be a minimum energy for the soliton, then
\begin{equation}
    E\ket{k} = H\ket{k} = X H^{\prime}\hat{A}\ket{0}.
\end{equation}
Operating with $X^{-1}$ from the left we get
\begin{equation} \label{trans}
    H^{\prime} \hat{A}\ket{0} = E\hat{A}\ket{0},
\end{equation}
where $\hat{A}\ket{0}$ is the lowest eigenvector corresponding to the transformed Hamiltonian. We get the lowest eigenstate (ground state) which solves eq. \eqref{trans}
\begin{equation}
    b_{k}\hat{A}\ket{0} = \pi_{0}\hat{A}\ket{0} = 0.
\end{equation}
We will work with the basis states by the eigenvectors $\ket{n}$ of the field operator $\phi(x)$ and let $n(x)$ be a real-valued function. The following eigenvalue equation defines the basis states 
\begin{equation}
    \phi(x) \ket{n} = n(x)\ket{n}.
\end{equation}
Let $\ket{\Psi}$ be expanded in terms of basis states of $n$. $\Psi$ is a complex-valued Schrodinger wave functional evaluated on the function $n$. This is an analogue to the wave function in Quantum mechanics. We will initially introduce the wave function and derive the wave function by calculating the expectation values of the basis states. The Schrodinger wave functional of the Sine-Gordon soliton state ($\Psi_{SG}(x)$) from \cite{main}is given by
\begin{equation}
    \Psi_{SG}(x) = exp\bigg(\frac{-1}{2}\int \frac{dk}{2\pi}(\phi_{k}-f_{-k})\omega_{k}(\phi_{-k}-f_{k})\bigg ),
\end{equation}
where $f_{k} = \int dx \phi^{cl}h_{k}(x)$. We shall go ahead and solve the wave functional to obtain the wave function.
$$\Psi_{SG} = exp \bigg(\frac{-1}{2}\int \frac{dk}{2\pi} \bigg[\phi_{k}\omega_{k}\phi_{-k}-\phi_{k}\omega_{k}f_{k}-f_{-k}\omega_{k}\phi_{-k}+f_{-k}\omega_{k}f_{k}\bigg ] \bigg ).$$
The expectation values can be calculated for the individual pieces which have the operators. We take the state $\ket{n}$ which is the eigenstate of the given system. The limit of $k$ is assumed to be between $-k_{0}$ and $k_{0}$. This gives us a finite result and so we choose to work with this assumption. 
$$\bra{n}exp\bigg(\frac{-1}{2}\int \frac{dk}{2\pi}(\phi_{k}\omega_{k}\phi_{-k})\bigg ) \ket{n} =  exp\bigg(-\frac{k_{0}}{4\pi}(n_{k}+n_{-k}+1)\bigg),\ n_{k},\ n_{-k}= 0,1,2..$$
\begin{multline} \label{fx}
    \bra{n}exp\bigg(\frac{-1}{2}\int \frac{dk}{2\pi}f_{-k}\omega_{k}f_{k} \bigg )\ket{n} =  \bra{n}exp\bigg(\frac{-1}{2}\int \frac{dk}{2\pi} \frac{16w_{k}^{2}}{\lambda(1+\frac{m^{2}}{k^{2}})}\int dx\bigg[arctan e^{mx}e^{-ikx}-\frac{im}{k}arctane^{mx}tanhmx\bigg]\\ \times \int dx^{\prime}\bigg[arctan e^{mx^{\prime}}e^{ikx^{\prime}}+\frac{im}{k}arctane^{mx^{\prime}}tanhmx^{\prime}\bigg]\bigg)\ket{n}.
\end{multline}
 We work by considering $x=x^{\prime}$ and after a tedious calculation, one finds the value of eq. \eqref{fx} as
\begin{equation}
     exp\bigg(\frac{-1}{2}\int \frac{dk}{2\pi}f_{-k}\omega_{k}f_{k} \bigg ) = exp\bigg(\frac{1}{2\pi}\bigg[\frac{m^{2}}{12k_{0}^{3}}+\frac{1}{4k_{0}}\bigg(\frac{\pi}{2}+mx\bigg)^{2}-\frac{\pi^{2}m^{2}}{24k_{0}^{3}}\bigg]\bigg).
\end{equation}
The expectation value of $exp\bigg[\frac{-1}{2}\int \frac{dk}{2\pi}\bigg(\phi_{k}\omega_{k}f_{k}-f_{-k}\omega_{k}\phi_{-k}\bigg)\bigg]$ is $0$. Therefore, the wave function of the Sine - Gordon soliton state $(\psi_{SG}(x))$ is
\begin{equation} \label{wavefunctionSG}
    \psi_{SG}(x)= exp\bigg(\frac{1}{2\pi}\bigg[\frac{m^{2}}{12k_{0}^{3}}+\frac{1}{4k_{0}}\bigg(\frac{\pi}{2}+mx\bigg)^{2}-\frac{\pi^{2}m^{2}}{24k_{0}^{3}}\bigg]-\bigg[\frac{k_{0}}{4\pi}(n_{k}+n_{-k}+1)\bigg]\bigg), n_{k},\ n_{-k}= 0,1,2.,
\end{equation}
where $k_{0}$ is a constant corresponding to the momentum in phase space. We can rewrite the previous equation as
\begin{equation}
    \psi_{SG}(x) = exp\bigg\{C+\frac{1}{8\pi k_{0}}\bigg(\frac{\pi}{2}+mx\bigg)^{2} \bigg\},
\end{equation}
where $C = \frac{1}{2\pi}\bigg[\frac{m^{2}}{12k_{0}^{3}}-\frac{\pi^{2}m^{2}}{24k_{0}^{3}}\bigg]-\bigg[\frac{k_{0}}{4\pi}(n_{k}+n_{-k}+1)\bigg],$  $n_{k},$ and $n_{-k}= 0,1,2$. Using eq. (\ref{eq:1}) we write the corresponding Wigner distribution for the Sine-Gordon soliton as
\begin{multline}
\mathcal{W}(x,k)= \frac{1}{\pi\hbar}\int^{\infty}_{-\infty} dy\Bigg\{exp\bigg(\frac{1}{2\pi}\bigg[\frac{m^{2}}{6k_{0}^{3}}+\frac{1}{4k_{0}}\Bigg[\bigg(\frac{\pi}{2}+m(x-y)\bigg)^{2}+\bigg(\frac{\pi}{2}+m(x+y)\bigg)^{2}\Bigg]-\frac{\pi^{2}m^{2}}{12k_{0}^{3}}\bigg]-\bigg[\frac{k_{0}}{4\pi}(n_{k}+n_{-k}+1)\bigg]\Bigg)\times \\exp\bigg(\frac{2iky}{\hbar}\bigg)  \bigg\}. 
\end{multline}
Upon integrating and simplifying the above relation we get the Wigner distribution of the Sine Gordon soliton state
\begin{equation}
    \mathcal{W}(x,k) = 2\pi \sqrt{\frac{k_{0}}{-m^{2}}} exp\Bigg\{\frac{1}{2\pi}\Bigg(\frac{m^{2}}{6k_{0}^{4}}+\frac{\pi^{2}}{8k_{0}}+\frac{m^{2}x^{2}+\pi mx}{2k_{0}}-\frac{\pi^{2}m^{2}}{12k_{0}^{3}}-\frac{k_{0}}{2}(n_{k}+n_{-k}+1)\Bigg)+\frac{\pi k_{0}}{m^{2}\hbar^{2}}k^{2}\Bigg\},
\end{equation}
where $n_{k}, n_{-k}= 0,1,2...$ Above equation can be written in more compact form using the process of completing the squares
\begin{equation}\label{eq:wigner}
    \mathcal{W}(x,k) = -2i\pi \sqrt{\frac{k_{0}}{m^{2}}}exp\Bigg\{\frac{1}{2\pi}\Bigg(A+\frac{1}{2k_{0}}\bigg[mx+\frac{\pi}{2}\bigg]^{2}\Bigg)+Bk^{2}\Bigg\},
\end{equation}
where $A= \frac{m^{2}}{6k_{0}^{4}}-\frac{\pi^{2}m^{2}}{12k_{0}^{3}}-\frac{k_{0}}{2}(n_{k}+n_{-k}+1), n_{k}, n_{-k}= 0, 1, 2.. $\\ \\
$B=\frac{\pi k_{0}}{m^{2}\hbar^{2}}.$\\ \\
Fig.(\ref{WDsineGordon}) shows the 3-dimensional plot of the Wigner distribution of Sine-Gordon soliton. We choose the values of $k_0= \hbar= m= 1$, and $n_k = n_{-k} = 0$ in order to plot fig.(\ref{WDsineGordon}). The magnitude of Wigner distribution is minimum at the origin and increases gradually on moving away from the origin. 
\pagebreak
\subsection{Charge and current density of Sine Gordon soliton}
The charge and current density can be obtained from the Wigner distribution. As we have already obtained the Wigner distributions for Sine -Gordon soliton in the previous section, we derive the charge $(\mathcal{Q}_{SG}(x))$  and current density for the same in this section. 
\begin{equation}
\label{eq:31}
    \mathcal{Q}_{SG}(x)=\frac{1}{\pi\hbar}\int^{\infty}_{-\infty} dk \int^{\infty}_{-\infty} dy \psi^{*}(x+y)\psi(x-y)e^{\frac{2iky}{\hbar}},
\end{equation}
and from eq. (\ref{eq:1}), we can write eq. (\ref{eq:31}) as
\begin{equation}
\label{eq:32}
    \mathcal{Q}_{SG}(x)=\int^{\infty}_{-\infty} dk \mathcal{W}(x,k). 
\end{equation}
Substituting the value of Wigner distribution from eq. (\ref{eq:wigner}) we get
\begin{equation}
\label{eq:34}
    \mathcal{Q}_{SG}(x)= \frac{-2i\pi \sqrt{\frac{k_{0}}{m^{2}}}}{\pi\hbar}\int^{\infty}_{-\infty} dk \Bigg[exp\Bigg\{\frac{1}{2\pi}\Bigg(A+\frac{1}{2k_{0}}\bigg[mx+\frac{\pi}{2}\bigg]^{2}\Bigg)+Bk^{2}\Bigg\}\Bigg],
\end{equation}
integrating and substituting the limits we get
\begin{equation}
\label{eq:36}
\mathcal{Q}_{SG}(x)= \frac{-2}{\hbar}\sqrt{\frac{\pi k_{0}}{m^{2}B}}exp\Bigg\{\frac{1}{2\pi}\Bigg(A+\frac{1}{2k_{0}}\bigg[mx+\frac{\pi}{2}\bigg]^{2}\Bigg)\Bigg\}.
\end{equation}
Upon substituting the value of $B=\frac{\pi k_{0}}{m^{2}\hbar^{2}} $,  we get the charge distribution as 
\begin{equation}
    \mathcal{Q}_{SG}(x)= -2exp\Bigg\{\frac{1}{2\pi}\Bigg(A+\frac{1}{2k_{0}}\bigg[mx+\frac{\pi}{2}\bigg]^{2}\Bigg)\Bigg\}.
\end{equation}
Therefore,
\begin{equation}
    |\mathcal{Q}_{SG}(x)| = 2exp\Bigg\{\frac{1}{2\pi}\Bigg(A+\frac{1}{2k_{0}}\bigg[mx+\frac{\pi}{2}\bigg]^{2}\Bigg)\Bigg\}\approx |\psi_{SG}(x)|^{2}.
\end{equation}
The computation of current density $\mathcal{J}_{SG}(x)$ from Wigner distribution is given by
\begin{equation}
\label{eq:37}
\mathcal{J}_{SG}(x)=\int^{\infty}_{-\infty} dk .k. \mathcal{W}(x,k).
\end{equation}
Substituting the values of Wigner distribution in eq. (\ref{eq:37}) we get
\begin{equation}
    \mathcal{J}_{SG}(x) = \frac{-2i\pi \sqrt{\frac{k_{0}}{m^{2}}}}{\pi\hbar}\int^{\infty}_{-\infty} dk \Bigg[ k\: exp\Bigg\{\frac{1}{2\pi}\Bigg(A+\frac{1}{2k_{0}}\bigg[mx+\frac{\pi}{2}\bigg]^{2}\Bigg)+Bk^{2}\Bigg\}\Bigg],
\end{equation}
integrating and substituting the limits we get
$$\mathcal{J}_{SG}(x) = 0. $$
\begin{figure}[hbt]
    \centering
    \includegraphics[scale=0.6]{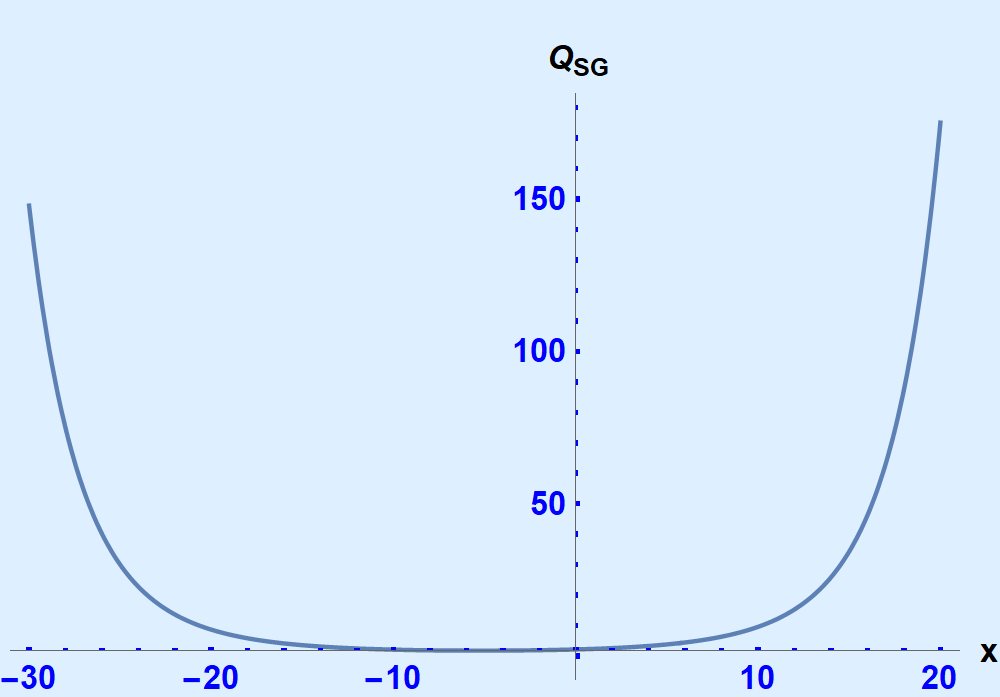}
    \caption{\label{chargesineGordon} Plot of charge density of Sine- Gordon soliton for $k_{0}\rightarrow 1, \hbar\rightarrow 1, n_{k} \rightarrow 0, n_{-k} \rightarrow 0, m\rightarrow 0.3.$  }
\end{figure} \\
The charge density of Sine-Gordon soliton as a function of position for a fixed value of $k_0=\hbar=1, ~m=0.3$, and $n_k=n_{-k}=0$ is shown in fig (\ref{chargesineGordon}). The charge distribution is minimum around the origin and increases exponentially as we move away from the origin.
\section{Wigner distribution of Kink soliton}\label{section:3}
We now shift our attention towards the kink soliton in this section. Kink soliton is the simplest topological soliton that arises in the theory of real scalar field in $1+1$ dimensional space-time. The Lagrangian density for the simple kink soliton is
$$ \mathcal{L}_{K}=\frac{1}{2}\partial^{\mu}\phi \partial_{\mu}\phi+ V(\phi),$$
where $V(\psi)$ represents the potential energy which can be given by
 $$V(\phi) = \frac{\lambda}{4}\phi^{2}(x)(\phi(x)-2a)^{2},$$
where $a=\frac{\mu}{\sqrt{\lambda}}$, the value of $\lambda$ depends on the particular system. The minimum of the potential occurs when $\frac{dV}{d\phi}=0$. The classical vacua occur at the minima of the potential. Therefore, they are at $\pm a$. $\phi$ is continuous and a transition region is observed between the two vacua. This is called the Domain wall \cite{soliton} and the symmetry is spontaneously broken upon the transformation, $\phi \to -\phi$ in this domain wall. Thus the kink soliton can be considered as a static solution of the field equations which is interpolating between two vacuum solutions. The classical function \cite{soliton} of the simplest kink soliton can be given by \cite{main}
\begin{equation}
\label{eq:5}
    \phi^{cl}_{K}(x) = \frac{m}{\sqrt{2\lambda}}(1+tanh\beta x),
\end{equation}
where $\beta = \frac{m}{2}$ and $m$ is the mass parameter. Since the soliton exists between its vacuum solutions we define the bound of the solitons between $-a$ and $+a$. The Hamiltonian density of the system can be given by
\begin{equation}
    \mathcal{H} = \frac{1}{2}:[\pi(x)]^{2}:+\frac{1}{2}:(\partial_{x}\phi(x))^{2}:-\frac{m^{2}}{\lambda} :(\mathrm{cos}(\sqrt{\lambda} \phi) - 1): .
\end{equation}
Normal ordered Hamiltonian of this system can be given by
\begin{equation}
    H = \int dx \mathcal{H} = \int dx \Bigg[\frac{1}{2}:[\pi(x)]^{2}:+\frac{1}{2}:(\partial_{x}\phi(x))^{2}: +  \frac{\lambda}{4}:\phi^{2}(x)(\phi(x)-2a)^{2}:\bigg].
\end{equation}
We follow the technique adopted in \cite{main} by defining a new Hamiltonian $H^{\prime}$ which is related to the original Hamiltonian through a similarity transformation. The new Hamiltonian $H^{\prime}$ is defined as
\begin{equation}
    H^{\prime} = X^{-1} H X, X=exp\bigg(-i\int dx \phi^{cl}\pi(x)\bigg),
\end{equation}
where $X$ is the translation operator which transforms the solution of the soliton. $H^{\prime}$ describes the oscillations in the soliton configuration. Since the Hamiltonian is normal ordered, regularization is not required. The transformed Hamiltonian $H^{\prime} = \mathcal{F}_{0}+\int dx \mathcal{H}^{\prime}$ where $\mathcal{F}_{0}$ represents the classical soliton energy, and
\begin{equation} \label{modifiedH}
    \mathcal{H}^{\prime} = \frac{:\pi(x)^{2}:}{2} + \frac{:(\partial_{x}\phi(x))^{2}}{2}:+\bigg(2\beta^{2}-3\beta^{2}sech^{2}(\beta x)\bigg ):\phi^{2}(x):,
\end{equation}
where $\beta = \frac{m}{2}$. The classical equations of motion have constant frequency solutions $g_{k}(x)$ which is parameterized by $k$, bound state solution $g_{B}(x)$ and classical bound state solution $g_{BO}(x)$ representing the soliton Goldstone mode
\begin{equation}
    g_{k}(x)= \frac{e^{-ikx}}{\sqrt{(1+\frac{\beta^{2}}{k^{2}})(1+\frac{4\beta^{2}}{k^{2}})}}\bigg(1+\frac{\beta^{2}}{k^{2}}(1-3tanh^{2}(\beta x)-3i\frac{\beta}{k}tanh(\beta x)\bigg ), g_{B}(x) = \frac{\sqrt{3\beta}}{2}sech^{2}(\beta x)
\end{equation}
and
\begin{equation}
    g_{BO} = -i\sqrt{\frac{3\beta}{2}}tanh(\beta x)sech(\beta x),
\end{equation}
with their respective frequencies
\begin{equation}
\omega_{BO}=\beta \sqrt{3}  , \omega_{B}= 0.
\end{equation}
By the property of normalization and orthogonality, we get
\begin{equation} \label{orthogonal}
\int dx |g_{BO}(x)|^{2} = 1 ,  g^{*}_{BO}(x) = -g_{BO}(x).
\end{equation}
It is convenient to decompose the field $\phi(x)$ to plane waves to obtain the Heisenberg operators $b_{k}$ in the ground state sector. The field $\phi(x)$ can be decomposed into the constant frequency solutions. We start with reviewing the quantisation of double well kink soliton solutions in their oscillator modes. We will follow the notations as per ref. \cite{main}. The field $\phi(x)$ and its conjugate $\pi(x)$ is defined as
$$\phi(x) = \phi_{0}g_{B}(x) + \phi_{BO}g_{BO}(x)+\int \frac{dk}{2\pi} \phi_{k}g_{k}(x),$$
$$\pi(x) = \pi_{0}g_{B}(x) + \pi_{BO}g_{BO}(x) + \int \frac{dk}{2\pi}\pi_{k}g_{k}(x),$$
where $\phi_{BO}= \frac{1}{\sqrt{2\omega_{BO}}}(b^{\dagger}_{BO}-b_{BO})$ and $\pi_{BO} = i\sqrt{\frac{\omega_{BO}}{2}}(b^{\dagger}_{BO}+b_{BO})$. By using the completeness relations we get,
$$\phi_{BO} = \int dx \phi(x)g^{*}_{BO}(x), \pi_{BO} = \int dx \pi(x)g^{*}_{BO}(x),$$
$$b^{\dagger}_{BO} = \sqrt{\frac{\omega_{BO}}{2}} \phi_{BO} - \frac{i}{\sqrt{2\omega_{k}}}\pi_{BO},$$
$$b_{BO} = -\sqrt{\frac{\omega_{BO}}{2}}\phi_{k}-\frac{i}{\sqrt{2\omega_{BO}}}\pi_{BO}.$$
The commutation relations are:
$$[\phi_{BO} ,pi_{BO}] = -i , [\phi^{*}_{BO} , \pi_{BO}]= i , [b_{BO} , b^{\dagger}_{BO}] = 1 .$$
We can find the value of the wave functional of the double well kink soliton by performing the procedure adopted in the previous section. The wave functional of the double well kink soliton ($\Psi_{K}$) for any state is given by \cite{main}
\begin{equation}\label{functionalkink}
    \Psi_{K} = exp\Bigg[-\frac{1}{2}\bigg(\phi_{BO}+ f_{BO}\bigg) \omega_{BO}\bigg(\phi^{*}_{BO}+f^{*}_{BO}\bigg)-\frac{1}{2}\int \frac{dk}{2\pi}\bigg(\phi_{k}-f_{-k}\bigg)\omega_{k}\bigg(\phi_{-k}-f_{k}\bigg)\bigg]
\end{equation}
where $f_{BO} = \int dx \phi^{cl}_{K}(x) g_{BO}(x),  f_{k}(x) = \int dx \phi^{cl}_{K}(x) g_{k}(x) $. Also, the soliton wave functional corresponding to the ground state ($\Psi^{(0)}_{K}$) is given by 
\begin{equation}
    \Psi^{(0)}_{K} = exp \bigg(-\frac{1}{2}\int \frac{dk}{2\pi} \phi_{k}\omega_{k}\phi_{-k}\bigg). 
\end{equation}
Now we can calculate the expectation value of each individual term which have the operators by taking the state $\ket{n}$ as the eigenstate of the system. We choose the limit of $k$ between $-k_{0}$ and $k_{0}$ as this choice gives us the finite result. Expanding eq. (\ref{functionalkink}),
\begin{multline}
      \Psi_{K} = exp\Bigg[-\frac{1}{2}\bigg(\phi_{BO}\omega_{BO}\phi^{*}_{BO}+\phi_{BO}\omega_{BO}f^{*}_{BO}+f_{BO}\omega_{BO}\phi^{*}_{BO}+f_{BO}\omega_{BO}f^{*}_{BO}\bigg)-\\ \frac{1}{2}\int \frac{dk}{2\pi}\bigg(\phi_{k}\omega_{k}\phi_{-k}-\phi_{k}\omega_{k}f_{k}-f_{-k}\omega_{k}\phi_{-k}+f_{-k}\omega_{k}f_{k}\bigg)\Bigg],
\end{multline}
$$\text{where}, \bra{n^{\prime}}\phi_{BO}\omega_{BO}\phi^{*}_{BO}\ket{n^{\prime}} = \frac{1}{2}(2n_{BO}^{\prime}+1),  n_{BO}^{\prime}=0,1,2., \text{and} $$
$$f_{BO}\omega_{BO}f^{*}_{BO} = \frac{3\sqrt{3}\beta^{3}}{2\lambda}\bigg[arctan(sinh\beta x)-sech\beta x tanh\beta x-2sech\beta x\bigg].$$
The expectation value of the remaining terms i.e. $exp\bigg[\frac{-1}{2}\int \frac{dk}{2\pi}\bigg(\phi_{k}\omega_{k}f_{k}-f_{-k}\omega_{k}\phi_{-k}\bigg)\bigg]$ and $exp\bigg[\frac{-1}{2}\bigg(f_{BO}\omega_{BO}\phi_{BO}^{*}+\phi_{BO}\omega_{BO}f_{BO}^{*}\bigg)\Bigg]$ are 0. 
$$\bra{n^{\prime\prime}}exp\bigg(-\frac{1}{2}\int \frac{dk}{2\pi}(\phi_{k}\omega_{k}\phi_{-k})\bigg ) \ket{n^{\prime\prime}} =  exp\bigg(-\frac{k_{0}}{4\pi}(n_{k}^{\prime\prime}+n_{-k}^{\prime\prime}+1)\bigg), n_{k}^{\prime\prime}, n_{-k}^{\prime\prime}= 0 ,1,2., \text{and}$$
\begin{multline*}
    \bra{n}exp\bigg(-\frac{1}{2}\int \frac{dk}{2\pi}f_{-k}\omega_{k}f_{k}\bigg)\ket{n} = \bra{n}exp\Bigg\{-\frac{1}{4\pi}\frac{2\beta^{2}}{\lambda}\Bigg[Ei(ik(x-x^{\prime}))(1+\beta(x+x^{\prime})+\beta^{2}xx^{\prime})+\frac{i4\beta^{2}(x-x^{\prime})}{k}\\ \bigg(-exp(ik(x-x^{\prime}))\bigg)+
    i(x-x^{\prime})E_{i}(ik(x-x^{\prime}))+\frac{\beta^{2}}{2k^{2}}\bigg(-exp(ik(x-x^{\prime}))(1+ik(x-x^{\prime}))\bigg)+k^{2}(x-x^{\prime})^{2}E_{i}(ik(x-x^{\prime}))\Bigg]\Bigg\}\ket{n}
\end{multline*}
where $E_{i}(ik(x-x^{\prime}))=\int \frac{dk}{k}exp(ik(x-x^{\prime}))$, which cannot be further reduced and is known as exponential integral. We choose $x=x^{\prime}$ that simplifies the above equation to
$$ exp\bigg(-\frac{1}{2}\int \frac{dk}{2\pi}f_{-k}\omega_{k}f_{k}\bigg) = exp\Bigg\{-\frac{1}{4\pi}\frac{2\beta^{2}}{\lambda}\Bigg[ln(k_{0})(1+2\beta x + \beta^{2}x^{2})+\frac{\beta^{2}}{k_{0}^{2}}\Bigg]\Bigg\}.  $$
Therefore the wave function for Kink soliton ($\psi_{K}(x)$) is
\begin{multline}
     \psi_{K}(x) = exp\Bigg\{-\frac{1}{4}(2n_{BO}^{\prime}+1)-\frac{3\sqrt{3}\beta^{3}}{4\lambda}\bigg[arctan(sinh\beta x)-sech\beta x tanh\beta x-2sech\beta x\bigg]-\\\frac{k_{0}}{4\pi}(n_{k}^{\prime\prime}+n_{-k}^{\prime\prime}+1)-\frac{1}{2\pi}\frac{\beta^{2}}{\lambda}\Bigg[ln(k_{0})(1+2\beta x + \beta^{2}x^{2})+\frac{\beta^{2}}{k_{0}^{2}}\Bigg]\Bigg\},
\end{multline}
where $n_{BO}^{\prime}=0, 1, 2, \text{and}$ $ n_{k}^{\prime\prime}, n_{-k}^{\prime\prime} = 0 , 1, 2,.$ The above equation can be rewritten in the following way
\begin{equation}
    \psi_{K}(x) = exp\Bigg\{D-\frac{3\sqrt{3}\beta^{3}}{4\lambda}\bigg[arctan(sinh\beta x)-sech\beta x tanh\beta x-2sech\beta x\bigg]-\frac{1}{2\pi}\frac{\beta^{2}}{\lambda}\Bigg[ln(k_{0})(1+2\beta x + \beta^{2}x^{2})+\frac{\beta^{2}}{k_{0}^{2}}\Bigg]\Bigg\}
\end{equation}
where $D= -\frac{1}{4}(2n_{BO}^{\prime}+1)-\frac{k_{0}}{4\pi}(n_{k}^{\prime\prime}+n_{-k}^{\prime\prime}+1), n_{BO}^{\prime}, n_{k}^{\prime\prime}, n_{-k}^{\prime\prime} = 0, 1 ,2, .$
Using eq. (\ref{eq:1}) we write the corresponding Wigner distribution for the Kink soliton as
\begin{multline}
\mathcal{W}(x,k)= \frac{1}{\pi\hbar}\int^{\infty}_{-\infty} dy\Bigg\{exp\bigg(D-\frac{3\sqrt{3}\beta^{3}}{4\lambda}\bigg[arctan(sinh\beta (x+y))-sech\beta (x+y) tanh\beta (x+y)-2sech\beta (x+y)\bigg]-\\ \frac{1}{2\pi}\frac{\beta^{2}}{\lambda}\Bigg[ln(k_{0})(1+2\beta (x+y) + \beta^{2}(x+y)^{2})+\frac{\beta^{2}}{k_{0}^{2}}\Bigg]\Bigg)\times exp\bigg(D-\frac{3\sqrt{3}\beta^{3}}{4\lambda}\bigg[arctan(sinh\beta (x-y))-sech\beta (x-y) tanh\beta (x-y)-\\ 2sech\beta (x-y)\bigg]-\frac{1}{2\pi}\frac{\beta^{2}}{\lambda}\Bigg[ln(k_{0})(1+2\beta (x-y) + \beta^{2}(x-y)^{2})+\frac{\beta^{2}}{k_{0}^{2}}\Bigg]\Bigg) exp\bigg(\frac{2iky}{\hbar}\bigg)\bigg\}. 
\end{multline}
Defining the function,\\ $f(x,y) = exp\bigg(D-\frac{3\sqrt{3}\beta^{3}}{4\lambda}\bigg[arctan(sinh\beta (x+y))-sech\beta (x+y) tanh\beta (x+y)-2sech\beta (x+y)\bigg]- \frac{1}{2\pi}\frac{\beta^{2}}{\lambda}\Bigg[ln(k_{0})(1+2\beta (x+y) + \beta^{2}(x+y)^{2})+\frac{\beta^{2}}{k_{0}^{2}}\Bigg]\Bigg)$, the Wigner distribution for Kink soliton can be written compactly as 
\begin{equation} \label{eq:wigner1kink}
    \mathcal{W}(x,k) = \frac{1}{\pi\hbar}\int^{\infty}_{-\infty} dy f(x,y) f(x,-y) exp\bigg(\frac{2iky}{\hbar}\bigg)   
\end{equation}
Wigner distribution of Kink soliton is calculated numerically and is plotted below in fig.(\ref{WDkink}). We choose $k_0=\hbar=m=\beta=\lambda=1$, and $n^\prime_{BO}=n^{\prime\prime}_{k}=n^{\prime\prime}_{-k}=0$ for both the plots of kink soliton. Ideally the integration over the $y$ should be from $-\infty$ to $\infty$, but we choose a cut off $-10$ to $10$ to perform the numerical integration for the plot in fig (\ref{WDkink}). The Wigner distribution of Kink soliton is symmetric in momentum space with its magnitude being maximum for zero momentum.
\begin{figure}[hbt]
    \centering
    \includegraphics[scale=0.8]{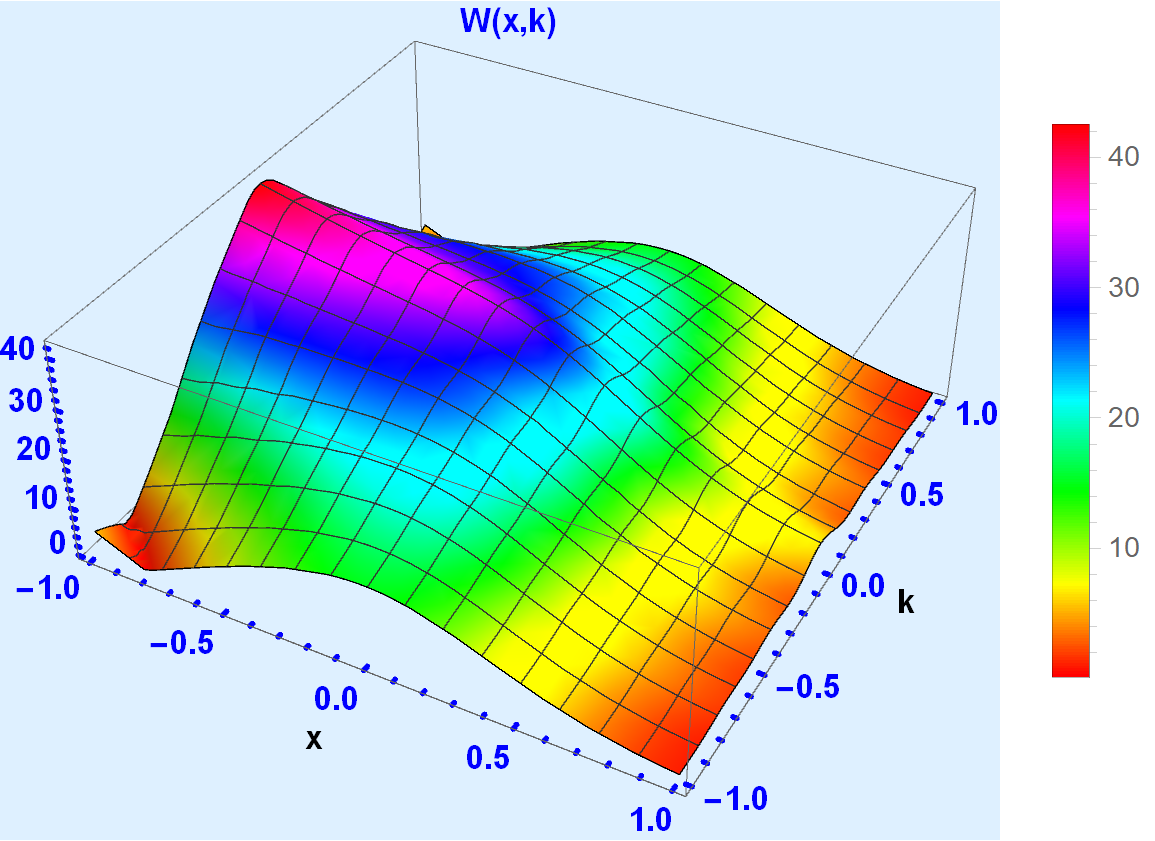}
    \caption{\label{WDkink} 3D plot of Wigner distribution for $k_{0}\rightarrow 1, \hbar\rightarrow 1, n_{BO}^{\prime} \rightarrow 0, n_{k}^{\prime\prime} \rightarrow 0, n_{-k}^{\prime\prime} \rightarrow 0 , m\rightarrow 1, \beta \rightarrow 1, \lambda \rightarrow 1.$}
\end{figure}

\newpage
\subsection{Charge and current density of Kink soliton}
The charge and current density can be obtained from the Wigner distribution. As we have already obtained
the Wigner distributions for Kink soliton in the previous section, we derive the charge and current density for the same in this section, numerically. The charge distribution $(Q_{K}(x))$ is obtained by
\begin{equation}
\label{eq:10}
    Q_{K}(x)=\frac{1}{h}\int_{-\infty}^{\infty} dp \int_{-\infty} ^{\infty} dy\bigg[\psi(x+y)\psi^{*}(x-y)e^{\frac{iPy}{\hbar}}\bigg],
\end{equation}
and from eq. (\ref{eq:1}) we can write eq. (\ref{eq:10}) as
\begin{equation}
\label{eq:11}
    Q_{K}(x)=\int_{-\infty}^{\infty} dk \mathcal{W}(x,k), 
\end{equation}
numerically evaluating the integration we find that the value of $Q_{K}(x)$ is approximately equal to $|\psi_{K}(x)|^{2}$. The result is plotted below in fig.(\ref{chargekink}). 
\begin{figure}[hbt]
    \centering
    \includegraphics[scale=0.6]{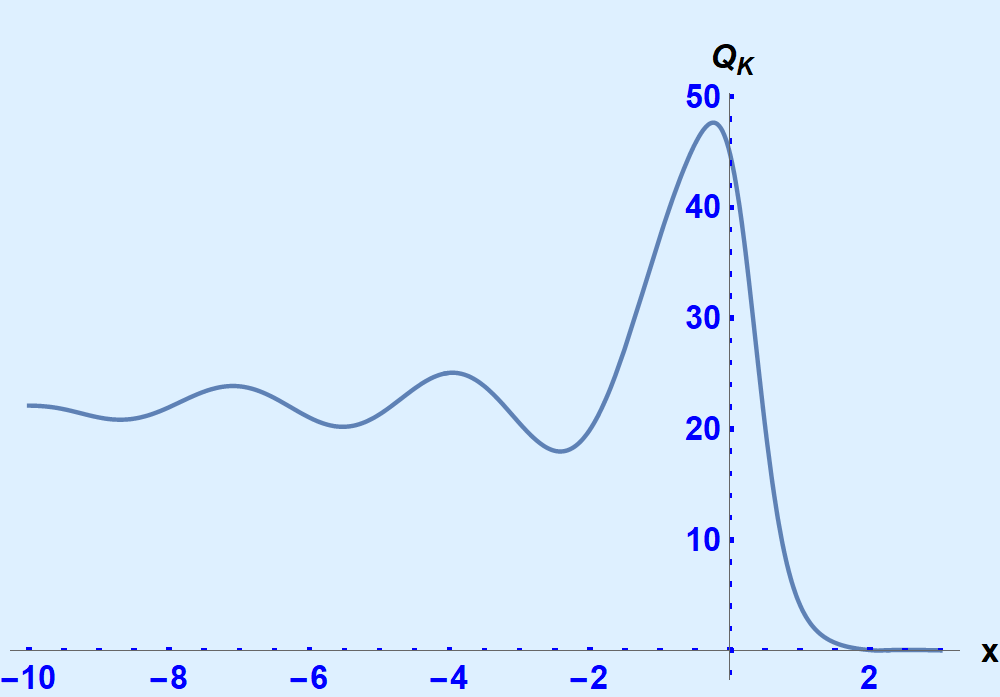}
    \caption{\label{chargekink} Plot of charge density of Kink soliton for $k_{0}\rightarrow 1, \hbar\rightarrow 1, n_{BO}^{\prime} \rightarrow 0, n_{k}^{\prime\prime} \rightarrow 0, n_{-k}^{\prime\prime} \rightarrow 0, m\rightarrow 1, \beta \rightarrow 1, \lambda \rightarrow 1.$  }
\end{figure}
The computation of current density $\mathcal{J}_{K}(x)$ from Wigner distribution is given by
\begin{equation}
\mathcal{J}_{K}(x)=\int^{\infty}_{-\infty} dk .k. \mathcal{W}(x,k),
\end{equation}
integrating and substituting the limits we get
$$\mathcal{J}_{K}(x) = 0. $$
Fig.(\ref{chargekink}) shows the charge distribution for the kink soliton. Again the integration over the $y$ should be from $-\infty$ to $\infty$, but we choose a cut off $-10$ to $10$ to perform the numerical integration for the plots in fig.(\ref{chargekink}). The charge density for the kink soliton is bounded for the positive $x$ axis, but is unbounded for the negative value of $x$. It shows peak near the origin and decrease on moving away from the origin on both side. On positive $x$ axis, charge density decreases sharply to zero, while on negative $x$ axis, it mimics a damped oscillation before becoming constant for large negative $x$.
\pagebreak

\section{Conclusion} \label{section:4}
Wigner distributions (quasi-probability distributions) play a significant role in formulating the phase space analogue of quantum mechanics. To calculate the Wigner distribution for a system, the quantum wave function for the system is needed. As the classical expression of solitons can not be used as wave functions, we evaluated the quantum wave function (Schrodinger wave-functional) for the Kink and Sine-Gordon soliton using the ``shifting of Hamiltonian" framework proposed in \cite{main} which makes the use of a classical expression of solitons. Using the Schrodinger wave-functional, we first obtained the analytical expression of Wigner distribution and then the charge and current density for both solitons. The Wigner distributions are analysed via plots and found to be symmetric in momentum space. The charge density in the position space is observed to be slightly shifted towards the negative of the origin which may be the result of choosing the ``Shifted Hamiltonian". The symmetric nature of Wigner distributions can also be seen from the 3-dimensional plots (Fig.(\ref{WDsineGordon}), Fig.(\ref{WDkink})) of the Wigner distributions. In addition, we derived the analytical expression of charge distributions and plotted them in  Fig.(\ref{chargesineGordon}) and Fig.(\ref{chargekink}). The current density was calculated and found zero for both Kink and Sine-Gordon solitons for the symmetric bounds. The Wigner distributions of the Kink soliton and the Sine-Gordon solitons are calculated in their respective {\it pure states}. These can be used to study the classical, semi-classical and quantum speed limit time \cite{deffner} which form the foundations of quantum computing. The quantum speed limit time gives us the value of the rate at which two quantum states are evolved \cite{deffner}. It can be calculated by calculating the value of quantum fidelity (QF) \cite{fidelity} - which measures the closeness of two states. Upon calculating it, we get the value of the quantum speed limit time as
\begin{equation}
\label{eq:41}
    \tau_{QSL} = \frac{1-F(t)}{\sqrt{2}}\frac{\hbar}{\Delta E},
\end{equation}
where $F(t)$ represents quantum fidelity. We can obtain quantum fidelity from Wigner distribution using the following expression
\begin{equation}
    F(t) = 2\pi\hbar \int dqdp W_{0}W_{t},
\end{equation}
where $W_{0}, W_{t}$ represent the Wigner distribution in the initial state and in the time-dependent state respectively. The quantum speed limit time \cite{deffner} which we have written above in eq. \eqref{eq:41} represents the Mandelstam-Tamm speed limit time \cite{tamm}. We have calculated the Wigner distributions of the kink and Sine-Gordon solitons in the current work. We will extend this work to calculate the same for the tunnelling instantons \cite{soliton} and also try to find the rate of evolution of two quantum states in the case of instantons.
\section{Acknowledgement} \label{section:5}
The authors are indebted to an anonymous referee for giving suggestions that have drastically improved the presentation of the manuscript
and also for pointing out the reference \cite{main}. Parts of the work were done while the author R.R. was at Sardar Vallabhbhai National Institute of Technology, Surat, India and at North Carolina State University, Raleigh, United States. The author R.R. would like to thank  Tiyasa Kar and Shaswat S. Tiwari for some helpful discussions. V.K.O. would like to thank the SVNIT Surat for the approval of the seed money project with the assigned project number 2021-22/DOP/05.
\appendix
\section{Computation of expectation values}
In this section we will discuss the process that was involved in computing the expectation values of various components in the wave functional. The Schrodinger wave functional of the Sine-Gordon soliton state ($\Psi_{SG}(x)$ is given by
$$ \Psi_{SG}(x) = exp\bigg(\frac{-1}{2}\int \frac{dk}{2\pi}(\phi_{k}-f_{-k})\omega_{k}(\phi_{-k}-f_{k})\bigg ), $$
where $f_{k} = \int dx \phi^{cl}h_{k}(x)$. We shall go ahead and solve the wave functional to obtain the wave function.
$$\Psi_{SG} = exp \bigg(\frac{-1}{2}\int \frac{dk}{2\pi} \bigg[\phi_{k}\omega_{k}\phi_{-k}-\phi_{k}\omega_{k}f_{k}-f_{-k}\omega_{k}\phi_{-k}+f_{-k}\omega_{k}f_{k}\bigg ] \bigg ).$$
We calculate the expectation values of the individual pieces which have the operators. We take the state $\ket{n}$ which is the eigenstate of the given system. The limit of $k$ is assumed to be between $-k_{0}$ and $k_{0}$. This gives us a finite result and so we choose to work with this assumption. 
$$\bra{n}exp\bigg(\frac{-1}{2}\int_{-k_{0}}^{k_{0}} \frac{dk}{2\pi}(\phi_{k}\omega_{k}\phi_{-k})\bigg ) \ket{n} =  exp\bigg(\frac{-1}{2}\int_{-k_{0}}^{k_{0}} \frac{dk}{2\pi}\bra{n}\phi_{k}\omega_{k}\phi_{-k}\ket{n}\bigg ), $$
$$ \bra{n}\phi_{k}\omega_{k}\phi_{-k}\ket{n} = \frac{1}{2}\bra{n}b_{k}^{\dagger}b_{-k}^{\dagger} + b_{k}^{\dagger}b_{k} + b_{-k}b_{-k}^{\dagger}+b_{-k}b_{k}\ket{n},  $$
$$ \bra{n}\phi_{k}\omega_{k}\phi_{-k}\ket{n} = \frac{1}{2}(n_{k}+n_{-k}+1)$$
since $\bra{n}b_{k}^{\dagger}b_{-k}^{\dagger}\ket{n}= \bra{n}b_{-k}b_{k}\ket{n} = 0$ and $\bra{n}b_{k}^{\dagger}b_{k}\ket{n} = n_{k}$, $\bra{n}b_{-k}b_{-k}^{\dagger}\ket{n} = n_{-k}+1$. Therefore, we get
$$\bra{n}exp\bigg(\frac{-1}{2}\int_{-k_{0}}^{k_{0}} \frac{dk}{2\pi}(\phi_{k}\omega_{k}\phi_{-k})\bigg ) \ket{n} =  exp\bigg(-\frac{k_{0}}{4\pi}(n_{k}+n_{-k}+1)\bigg), n_{k}\\, n_{-k}= 0, 1, 2., $$
$$\bra{n} exp\bigg[\frac{-1}{2}\int \frac{dk}{2\pi}\bigg(\phi_{k}\omega_{k}f_{k}-f_{-k}\omega_{k}\phi_{-k}\bigg)\bigg]\ket{n} = 0. $$
This is due the fact $\bra{n}\phi_{k}\omega_{k}f_{k}\ket{n} =\frac{\omega_{k}}{\sqrt{2\omega_{k}}}f_{k} \bra{n}(b_{k}^{\dagger}+b_{-k})\ket{n} = 0  $ and $\bra{n}f_{-k}\omega_{k}\phi_{-k}\ket{n} = \frac{\omega_{k}}{\sqrt{2\omega_{k}}}f_{-k}\bra{n}b_{k}+b_{-k}^{\dagger}\ket{n}= 0. $
\begin{multline*}
     \bra{n}exp\bigg(\frac{-1}{2}\int \frac{dk}{2\pi}f_{-k}\omega_{k}f_{k} \bigg )\ket{n} =  \bra{n}exp\bigg(\frac{-1}{2}\int \frac{dk}{2\pi} \frac{16w_{k}^{2}}{\lambda(1+\frac{m^{2}}{k^{2}})}\int dx\bigg[arctan e^{mx}e^{-ikx}-\frac{im}{k}arctane^{mx}tanhmx\bigg]\\ \times \int dx^{\prime}\bigg[arctan e^{mx^{\prime}}e^{ikx^{\prime}}+\frac{im}{k}arctane^{mx^{\prime}}tanhmx^{\prime}\bigg]\bigg)\ket{n}.
\end{multline*}
After a tedious calculation one finds
$$ exp\bigg(\frac{-1}{2}\int \frac{dk}{2\pi}f_{-k}\omega_{k}f_{k} \bigg ) = exp\bigg(\frac{1}{2\pi}\bigg[\frac{m^{2}}{12k_{0}^{3}}+\frac{1}{4k_{0}}\bigg(\frac{\pi}{2}+mx\bigg)^{2}-\frac{\pi^{2}m^{2}}{24k_{0}^{3}}\bigg]\bigg).$$
Therefore the wave function of the Sine- Gordon soliton state $(\psi_{SG}(x))$ is given by
$$ \bra{n}\psi_{SG}(x)\ket{n}= \bra{n}exp\bigg\{\frac{1}{2\pi}\bigg[\frac{m^{2}}{12k_{0}^{3}}+\frac{1}{4k_{0}}\bigg(\frac{\pi}{2}+mx\bigg)^{2}-\frac{\pi^{2}m^{2}}{24k_{0}^{3}}\bigg]-\bigg[\frac{k_{0}}{4\pi}(n_{k}+n_{-k}+1)\bigg]\bigg\}\ \ket{n}, n_{k}, n_{-k}= 0 ,1 ,2.. $$
The wave functional of the double well kink soliton ($\Psi_{K}$) for any state is given by
$$   \Psi_{K} = exp\Bigg[-\frac{1}{2}\bigg(\phi_{BO}+ f_{BO}\bigg) \omega_{BO}\bigg(\phi^{*}_{BO}+f^{*}_{BO}\bigg)-\frac{1}{2}\int \frac{dk}{2\pi}\bigg(\phi_{k}-f_{-k}\bigg)\omega_{k}\bigg(\phi_{-k}-f_{k}\bigg)\bigg], $$
expanding the previous equation we get
\begin{multline*}
      \Psi_{K} = exp\Bigg[-\frac{1}{2}\bigg(\phi_{BO}\omega_{BO}\phi^{*}_{BO}+\phi_{BO}\omega_{BO}f^{*}_{BO}+f_{BO}\omega_{BO}\phi^{*}_{BO}+f_{BO}\omega_{BO}f^{*}_{BO}\bigg)-\\ \frac{1}{2}\int \frac{dk}{2\pi}\bigg(\phi_{k}\omega_{k}\phi_{-k}-\phi_{k}\omega_{k}f_{k}-f_{-k}\omega_{k}\phi_{-k}+f_{-k}\omega_{k}f_{k}\bigg)\Bigg].
\end{multline*}
We calculate the expectation values of the individual pieces which have operators. The limit of $k$ is assumed to be between $-k_{0}$ and $k_{0}$. This gives us a finite result and so we choose to work with this assumption.
$$\bra{n^{\prime}}exp\bigg[-\frac{1}{2}\bigg(\phi_{BO}\omega_{BO}\phi^{*}_{BO}\bigg)\bigg] \ket{n^{\prime}} = exp\bigg[-\frac{1}{2}\bra{n^{\prime}}\bigg(b_{BO}^{\dagger}b_{BO}^{\dagger} + b_{BO}^{\dagger}b_{BO} + b_{BO}b_{BO}^{\dagger}+b_{BO}b_{BO}\ket{n^{\prime}}\bigg)\bigg],  $$
$$\bra{n^{\prime}}b_{BO}^{\dagger}b_{BO}^{\dagger} + b_{BO}^{\dagger}b_{BO} + b_{BO}b_{BO}^{\dagger}+b_{BO}b_{BO}\ket{n^{\prime}} = \frac{1}{2}(2n_{BO}^{\prime}+1),  $$
since $\bra{n^{\prime}}b_{BO}^{\dagger}b_{BO}^{\dagger}\ket{n^{
\prime}}= \bra{n^{\prime}}b_{BO}b_{BO}\ket{n^{\prime}} = 0$ and $\bra{n^{\prime}}b_{BO}^{\dagger}b_{BO}\ket{n^{\prime}} = n^{\prime}_{BO}$, $\bra{n^{\prime}}b_{BO}b_{BO}^{\dagger}\ket{n^{\prime}} = n^{\prime}_{BO}+1$. Therefore
$$\bra{n^{\prime}}exp\bigg[-\frac{1}{2}\bigg(\phi_{BO}\omega_{BO}\phi^{*}_{BO}\bigg)\bigg] \ket{n^{\prime}} = exp\bigg[-\frac{1}{4}(2n_{BO}^{\prime}+1)\bigg],  $$
$$\bra{n^{\prime}} exp\bigg[\frac{-1}{2}\bigg(\phi_{BO}\omega_{BO}f^{*}_{BO}+f_{BO}\omega_{BO}\phi^{*}_{BO}\bigg)\bigg]\ket{n^{\prime}} = 0. $$
This is due the fact, $\bra{n^{\prime}}\phi_{BO}\omega_{BO}f^{*}_{BO}\ket{n^{\prime}} =\frac{\omega_{BO}}{\sqrt{2\omega_{BO}}}f^{*}_{BO} \bra{n^{\prime}}(b_{BO}^{\dagger}+b_{BO})\ket{n^{\prime}} = 0  $ and $\bra{n^{\prime}}f_{BO}\omega_{BO}\phi^{*}_{BO}\ket{n^{\prime}} = \frac{\omega_{BO}}{\sqrt{2\omega_{BO}}}f_{BO}\bra{n^{\prime}}b_{BO}+b_{BO}^{\dagger}\ket{n^{\prime}}= 0 $. 
$$f_{BO}\omega_{BO}f^{*}_{BO} = \frac{3\sqrt{3}\beta^{3}}{2\lambda}\bigg[arctan(sinh\beta x)-sech\beta x tanh\beta x-2sech\beta x\bigg],$$
thus,
$$exp-\frac{1}{2}\bigg[f_{BO}\omega_{BO}f^{*}_{BO}\bigg] = exp\bigg\{ -\frac{3\sqrt{3}\beta^{3}}{4\lambda}\bigg[arctan(sinh\beta x)-sech\beta x tanh\beta x-2sech\beta x\bigg]\bigg\}. $$
Now we find the expectation values of second part of the expression.
$$\bra{n^{\prime\prime}}exp\bigg(\frac{-1}{2}\int_{-k_{0}}^{k_{0}} \frac{dk}{2\pi}(\phi_{k}\omega_{k}\phi_{-k})\bigg ) \ket{n^{\prime\prime}} =  exp\bigg(\frac{-1}{2}\int_{-k_{0}}^{k_{0}} \frac{dk}{2\pi}\bra{n^{\prime\prime}}\phi_{k}\omega_{k}\phi_{-k}\ket{n^{\prime\prime}}\bigg ), $$
$$ \bra{n^{\prime\prime}}\phi_{k}\omega_{k}\phi_{-k}\ket{n^{\prime\prime}} = \frac{1}{2}\bra{n^{\prime\prime}}b_{k}^{\dagger}b_{-k}^{\dagger} + b_{k}^{\dagger}b_{k} + b_{-k}b_{-k}^{\dagger}+b_{-k}b_{k}\ket{n^{\prime\prime}},  $$
$$ \bra{n^{\prime\prime}}\phi_{k}\omega_{k}\phi_{-k}\ket{n^{\prime\prime}} = \frac{1}{2}(n^{\prime\prime}_{k}+n^{\prime\prime}_{-k}+1)$$
Since $\bra{n^{\prime\prime}}b_{k}^{\dagger}b_{-k}^{\dagger}\ket{n^{\prime\prime}}= \bra{n^{\prime\prime}}b_{-k}b_{k}\ket{n^{\prime\prime}} = 0$ and $\bra{n^{\prime\prime}}b_{k}^{\dagger}b_{k}\ket{n^{\prime\prime}} = n^{\prime\prime}_{k}$, $\bra{n^{\prime\prime}}b_{-k}b_{-k}^{\dagger}\ket{n^{\prime\prime}} = n^{\prime\prime}_{-k}+1$. Therefore we get
$$\bra{n^{\prime\prime}}exp\bigg(\frac{-1}{2}\int_{-k_{0}}^{k_{0}} \frac{dk}{2\pi}(\phi_{k}\omega_{k}\phi_{-k})\bigg ) \ket{n^{\prime\prime}} =  exp\bigg(-\frac{k_{0}}{4\pi}(n^{\prime\prime}_{k}+n^{\prime\prime}_{-k}+1)\bigg), n^{\prime\prime}_{k}, n^{\prime\prime}_{-k}= 0, 1, 2. , $$
$$\bra{n^{\prime\prime}} exp\bigg[\frac{-1}{2}\int \frac{dk}{2\pi}\bigg(\phi_{k}\omega_{k}f_{k}-f_{-k}\omega_{k}\phi_{-k}\bigg)\bigg]\ket{n^{\prime\prime}} = 0. $$
This is due the fact $\bra{n^{\prime\prime}}\phi_{k}\omega_{k}f_{k}\ket{n^{\prime\prime}} =\frac{\omega_{k}}{\sqrt{2\omega_{k}}}f_{k} \bra{n^{\prime\prime}}(b_{k}^{\dagger}+b_{-k})\ket{n^{\prime\prime}} = 0  $ and $\bra{n^{\prime\prime}}f_{-k}\omega_{k}\phi_{-k}\ket{n^{\prime\prime}} = \frac{\omega_{k}}{\sqrt{2\omega_{k}}}f_{-k}\bra{n^{\prime\prime}}b_{k}+b_{-k}^{\dagger}\ket{n^{\prime\prime}}= 0 $. Finally we calculate the expectation value of $f_{-k}\omega_{k}f_{k} $.
\begin{multline*}
    f_{-k}\omega_{k}f_{k} = \int dx \frac{m}{\sqrt{2\lambda}}\bigg(1+ tanh\bigg(\frac{mx}{2}\bigg)\bigg) \frac{exp(ikx)}{\sqrt{(1+\frac{\beta^{2}}{k^{2}})(1+\frac{4\beta^{2}}{k^{2}})}}\Bigg(1+\frac{\beta^{2}}{k^{2}}(1-3tanh^{2}\beta x)+\frac{3i\beta}{k}tanh\beta x\Bigg) \sqrt{m^{2}+k^{2}} \\ \int dx^{\prime} \frac{m}{\sqrt{2\lambda}}\bigg(1+ tanh\bigg(\frac{mx^{\prime}}{2}\bigg)\bigg) \frac{exp(-ikx^{\prime})}{\sqrt{(1+\frac{\beta^{2}}{k^{2}})(1+\frac{4\beta^{2}}{k^{2}})}}\Bigg(1+\frac{\beta^{2}}{k^{2}}\bigg(1-3tanh^{2}\beta x^{\prime}-3i\frac{\beta}{k}tanh\beta x^{\prime}\bigg)\Bigg),
\end{multline*}
$$ f_{-k}\omega_{k}f_{k} = \frac{\sqrt{m^{2}+k^{2}}m^{2}}{(1+\frac{\beta^{2}}{k^{2}})(1+\frac{4\beta^{2}}{k^{2}})} I_{1}I_{2} $$
where $I_{1} =  \int dx \bigg(1+ tanh\beta x\bigg)exp(ikx)\Bigg(1+\frac{\beta^{2}}{k^{2}}(1-3tanh^{2}\beta x)+\frac{3i\beta}{k}tanh\beta x\Bigg), $\\ 
$I_{2} = \int dx^{\prime}  \bigg(1+ tanh\beta x^{\prime}\bigg)exp(-ikx^{\prime}) \Bigg(1+\frac{\beta^{2}}{k^{2}}\bigg(1-3tanh^{2}\beta x^{\prime}-3i\frac{\beta}{k}tanh\beta x^{\prime}\bigg)\Bigg) $.
We evaluate $I_{1}$ and $I_{2}$ separately.
$$I_{1} = \int dx exp(ikx)\bigg(1+\frac{\beta^{2}}{k^{2}}(1-3\beta^{2}x^{2})+\frac{3i\beta}{k}(\beta x)+\beta x+\frac{\beta^{2}}{k^{2}}(1-3\beta^{2}x^{2})+\frac{3i\beta}{k}\beta^{2}x^{2}\bigg)$$
$$I_{1} = \bigg(3\frac{\beta^{2}}{k^{2}}x-\frac{\beta}{k^{2}}\bigg)exp(ikx)-i\bigg(\frac{1}{k}+\frac{\beta^{2}}{k^{2}}-\frac{3\beta^{2}}{k^{3}}+\frac{\beta x}{k}\bigg) exp(ikx),$$
$$I_{1} = \bigg(3\frac{\beta^{2}}{k^{2}}x-\frac{\beta}{k^{2}}\bigg)exp(ikx)-i\bigg(\frac{1}{k}-\frac{2\beta^{2}}{k^{3}}+\frac{\beta x}{k}\bigg) exp(ikx),  $$
$$I_{2} = \bigg(\frac{3\beta^{2}x^{\prime}}{k^{2}}-\frac{\beta}{k^{2}}\bigg)exp(-ikx^{\prime})-i\bigg(\frac{1}{k}-\frac{2\beta^{2}}{k^{3}}+\frac{\beta x^{\prime}}{k}\bigg) exp(-ikx^{\prime}).$$
Therefore,
\begin{multline*}
    f_{-k}\omega_{k}f_{k} = \frac{\sqrt{m^{2}+k^{2}}m^{2}}{2\lambda(1+\frac{\beta^{2}}{k^{2}})(1+\frac{4\beta^{2}}{k^{2}})}\Bigg[3\frac{\beta^{2}}{k^{2}}x-\frac{\beta}{k^{2}}-i\bigg(\frac{1}{k}-\frac{2\beta^{2}}{k^{3}}+\frac{\beta x}{k}\bigg)\Bigg]\Bigg[\frac{3\beta^{2}x^{\prime}}{k^{2}}-\frac{\beta}{k^{2}}-i\bigg(\frac{1}{k}-\frac{2\beta^{2}}{k^{3}}+\frac{\beta x^{\prime}}{k}\bigg)\Bigg]exp(ik(x-x^{\prime})).     
\end{multline*}
We work till $O(\beta^{2})$
$$\int dk  f_{-k}\omega_{k}f_{k} = \int dk \frac{\sqrt{m^{2}+k^{2}}m^{2}exp(ik(x-x^{\prime}))}{2\lambda(1+\frac{\beta^{2}}{k^{2}})(1+\frac{4\beta^{2}}{k^{2}})}\Bigg[\frac{1}{k^{2}}\bigg(1+\beta(x+x^{\prime})+\beta^{2}xx^{\prime}\bigg) + \frac{i}{k^{3}}\bigg(4\beta^{2}(x-x^{\prime})\bigg)+\frac{\beta^{2}}{k^{4}}\Bigg],  $$ 
\begin{multline*}
    exp\bigg(-\frac{1}{2}\int \frac{dk}{2\pi}f_{-k}\omega_{k}f_{k}\bigg) = exp\Bigg\{-\frac{1}{4\pi}\frac{2\beta^{2}}{\lambda}\Bigg[Ei(ik(x-x^{\prime}))(1+\beta(x+x^{\prime})+\beta^{2}xx^{\prime})+\frac{i4\beta^{2}(x-x^{\prime})}{k}\bigg(-exp(ik(x-x^{\prime}))\bigg)+\\
    i(x-x^{\prime})E_{i}(ik(x-x^{\prime}))+\frac{\beta^{2}}{2k^{2}}\bigg(-exp(ik(x-x^{\prime}))(1+ik(x-x^{\prime}))\bigg)+k^{2}(x-x^{\prime})^{2}E_{i}(ik(x-x^{\prime}))\Bigg]\Bigg\}
\end{multline*}
where $Ei(ik(x-x^{\prime}))=\int \frac{dk}{k}exp(ik(x-x^{\prime}))$, which cannot be further reduced and it is called the exponential integral. We work in the domain by considering $x=x^{\prime}$, thus we get
$$ exp\bigg(-\frac{1}{2}\int \frac{dk}{2\pi}f_{-k}\omega_{k}f_{k}\bigg) = exp\Bigg\{-\frac{1}{4\pi}\frac{2\beta^{2}}{\lambda}\Bigg[ln(k_{0})(1+2\beta x + \beta^{2}x^{2})+\frac{\beta^{2}}{k_{0}^{2}}\Bigg]\Bigg\}.  $$
Therefore the wave function for Kink soliton ($\psi_{K}(x)$) can be given by
\begin{multline*}
     \psi_{K}(x) = exp\Bigg\{-\frac{1}{4}(2n_{BO}^{\prime}+1)-\frac{3\sqrt{3}\beta^{3}}{4\lambda}\bigg[arctan(sinh\beta x)-sech\beta x tanh\beta x-2sech\beta x\bigg]-\\\frac{k_{0}}{4\pi}(n_{k}^{\prime\prime}+n_{-k}^{\prime\prime}+1)-\frac{1}{2\pi}\frac{\beta^{2}}{\lambda}\Bigg[ln(k_{0})(1+2\beta x + \beta^{2}x^{2})+\frac{\beta^{2}}{k_{0}^{2}}\Bigg]\Bigg\},
\end{multline*}
where $n_{BO}^{\prime}=0, 1, 2, $, $n_{k}^{\prime\prime}, n_{-k}^{\prime\prime} = 0,1,2..$

\end{document}